\documentclass[journal]{IEEEtran}

\usepackage[dvips]{graphicx}
\graphicspath{{./}}
\DeclareGraphicsExtensions{.eps}
\usepackage[cmex10]{amsmath}
\begin{document}

\title{Channel Independent Precoder for OFDM-based Systems over Fading Channels}

\author{Jorge~Ort\' in,~\IEEEmembership{Student~Member,~IEEE,}
       Paloma~Garc\' ia,~Fernando~Guti\' errez~and~Antonio~Valdovinos%
\thanks{This work has been suported by the Spanish Government (FPU grant and Project TEC 2008-06684-C03-02/TEC), Gobierno de Arag\'on (Project PI003/08 and WALQA Technology Park) and the European IST Project EUWB.

The authors are with the Arag\'on Institute for Engineering Research (I3A), University of Zaragoza, Zaragoza, 50018, Spain (e-mails: jortin@unizar.es, paloma@unizar.es, ferguso@unizar.es, toni@unizar.es). 
}}

\maketitle

\begin{abstract}
In this paper we propose an independent channel precoder for orthogonal frequency division multiplexing (OFDM) systems over fading channels. The design of the precoder is based on the information redistribution of the input modulated symbols amongst the output precoded symbols. The proposed precoder decreases the variance of the instantaneous noise power at the receiver produced by the channel variability. The employment of an interleaver together with a precoding matrix whose size does not depend on the number of data carriers in an OFDM symbol allows different configurations of time-frequency diversity which can be easily adapted to the channel conditions. The precoder is evaluated with a modified Zero Forcing (ZF) equalizer whose maximum gain is constrained by means of a clipping factor. Thus, the clipping factor limits the noise power transfer in the receiver deprecoding block in low SNR conditions.
\end{abstract}


\begin{figure*}[!t]
\centering
\includegraphics[width=6.5in, clip=true]{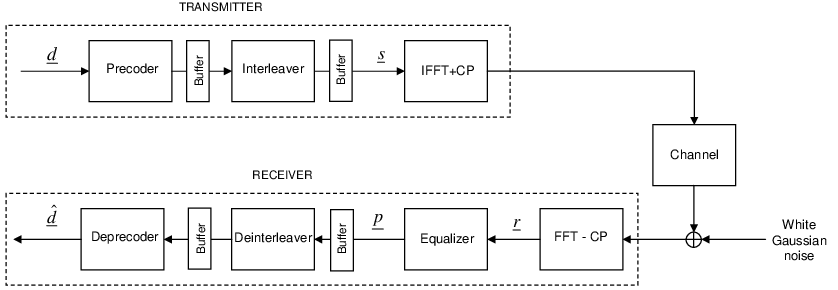}
\caption{System Block Diagram}
\label{fig:blocks}
\end{figure*}

\section{Introduction}
\IEEEPARstart{O}{rthogonal} frequency division multiplexing (OFDM) is a multicarrier modulation technology which has been selected as the preferred modulation for the new high speed communications systems. Multicarrier-based modulations are employed in a wide range of wired and wireless scenarios, such as xDSL for the subscriber local loop access \cite{general:xDSL}, digital radio broadcasting (DAB) \cite{general:DAB}, digital terrestrial television broadcasting (DVB-T) \cite{general:DVB-T} and its version for mobile handsets (DVB-H) \cite{general:DVB-H}, wireless local area networks \cite{general:802.11} and wireless metropolitan access networks \cite{general:802.16}. OFDM is also one the selected technologies for the evolution of the 3G mobile networks \cite{general:LTE}.

OFDM modulation divides the user data flow into several streams which are modulated in different carriers at a low symbol rate. The use of a longer symbol period together with a guard interval mitigates the effect of the intersymbolic interference (ISI). Furthermore, the choice of suitable subcarrier spacing causes that each subcarrier exhibits a flat fading channel, simplifying the equalization at the receiver. Nevertheless, under severe conditions of propagation, such as the ones typical of mobile channels, some of the subcarriers could suffer high attenuation causing a raise in the raw average bit error rate (BER). While the application of forward error correcting codes decreases the effective BER at the expense of a throughput loss, the use of precoding schemes can improve the overall BER without the introduction of redundant information \cite{precoding:Wang1}.

The use of precoders aims at the compensation of the impairments produced by the channel not only at the receiver but also at the transmitter. Most of the proposed precoders for OFDM systems are based on the total \cite{precoding:Ding1, precoding:Scaglione1} or partial \cite{precoding:Liu1, precoding:Rong1} knowledge of the transmission channel at the transmitter. This knowledge is unavailable in broadcasting systems where there is not a feedback link between transmitter and receiver and difficult to achieve in cellular systems due to the high variability of the channel and the resource consumption produced by the channel information feedback. In these cases, the development of channel independent precoders could be of paramount importance in order to improve the BER without any prior channel information at the transmitter.

In \cite{precoding:Lin1} an independent channel precoder is proposed and it is demonstrated that depending on the precoding matrix selected, it is equivalent to a single carrier system. Likewise, if the matrix employed to perform the precoding is a Hadamard matrix, the result is a orthogonal frequency division multiplexing - code division multiplexing (OFDM-CDM) system \cite{precoding:Kaiser}, \cite{precoding:book_precoding}. In this sense, OFDM-CDM is a special case of the precoder described in \cite{precoding:Lin1}. Nevertheless, the typical configuration of those systems constrains the precoder block size to the number of carriers in an OFDM symbol and the symbols corresponding to the same precoding block are mapped to consecutive carriers in the same OFDM symbol. 

These systems are similar to a multicarrier - code division multiple access (MC-CDMA) system \cite{MC-CDMA:Yee1}, \cite{MC-CDMA:Chin1}, where the processing of the precoder input data is the same as the frequency spreading performed to the different users' data flows in a MC-CDMA system. From this point of view, any proposed solution in order to improve the performance of a MC-CDMA system can be adapted to be used as an OFDM channel independent precoder.

In that sense, the proposed system in \cite{MC-CDMA:Zheng1} could be applied to precode information data in an OFDM system. This system spreads the user data not only in the frequency plane but also in the time domain, generating a 2D spread signal. Although this system improves the overall performance because of the increase in the diversity of the transmitted signal, it does not completely use the statistical properties of the channel since the spreading is done in contiguous symbols, so it is not ensured the independence of the channel amongst the slots corresponding to the same spread symbol. The spreading factor is also equal to the number of available subcarriers and it is performed a Maximum Ratio Combining (MRC) at the receiver, which is suitable for MC-CDMA but can distort the signal amplitude if applied to the carriers of an OFDM symbol. In that sense, MRC only ensures phase equalization but not amplitude equalization, so it is not suitable for multilevel modulated carriers.

In this paper, we propose a channel independent precoder for OFDM systems with variable block size, which can be different from the size of the fast Fourier transform (FFT) employed in the system. This precoder is followed by a time-frequency interleaver which will ensure that the data corresponding to the same precoded block are moved apart a distance higher than the channel coherence bandwidth/time. The equalization at the receiver is performed with a clipped-modified Zero Forcing (ZF) filter whose maximum gain is limited in order to restrict the noise amplification. The clipping factor is adjusted to maximize the BER at the inverse precoder output. Higher clipping factors will raise the noise power transfer amongst the output data while lower clipping factors will distort the signal even in noise absence. The performance of the proposed system is evaluated in realistic radio channels which present multipath frequency-selective fadings and are time variant.

This paper is organized as follows. Section II gives an overall description of the proposed precoding system. In section III an analytical approach of the system shows in detail the operation of the proposed precoding scheme. Section IV analyses the effect of the receiver equalizer on the precoder performance. In section V simulation results are presented and discussed. Finally, the conclusions are summarized in section VI.

\section{Precoding System Description}

The block diagram of the proposed system is shown in Fig. \ref{fig:blocks}. The precoder at the transmitter is located between the mapping of the binary data to modulated symbols (i. e. QPSK) and the transmission block, which computes the IFFT and adds the cyclic prefix (CP). The location of the deprecoding block at the receiver is after the equalizer and before the demodulation block. The modulated symbols are partitioned into blocks of size $N$, with $N$ the precoder block size. Each input vector $\textbf{d}$ of modulated symbols is then multiplied by the Hadamard matrix $\textbf{H}$ of size $N\times{N}$ to obtain the precoded vector $\textbf{s}$. This matrix meets the requirements for the information redistribution of the input data as explained in the next paragraph. After this multiplication, several precoded blocks are interleaved in order to ensure channel independence amongst all the symbols belonging the same precoded block. The output is then partitioned into blocks fitting the number of user data carriers of an OFDM symbol. Hence, the precoder block size could be different from the size of the IFFT performed on transmission. 

The objective of precoding is to redistribute the information of each modulated symbol into all the output symbols of the corresponding precoder block. This fact allows the span of a modulated symbol into several OFDM carriers, increasing the transmission diversity and averaging the channel attenuations suffered by the modulated data in a precoded block. In order to make this redistribution, it is needed that all the elements in the precoding matrix have the same magnitude so that every output symbol has the same amount of information of every input data. This magnitude has to be equal to $1/\sqrt{N}$ to keep the power at the precoder output. Additionally, the precoding matrix has to be non-singular so that the recovery of the original data is ensured. 

The normalised Hadamard matrix fulfils the conditions stated in the previous paragraph and holds additional interesting properties. The Hadamard matrix is a square matrix whose elements are either $+1$ or $-1$ and whose rows are mutually orthogonal, therefore its inverse corresponds to its transpose. The Hadamard matrix is also easy to operate with in a signal processor since it only requires additions and substractions being the computational load low.

The received vector after the FFT calculation and the CP discarding can be  expressed as:

\begin{equation}
\textbf{r} = \textbf{H} \, \textbf{s} + \textbf{n} = \textbf{H} \, \textbf{P} \, \textbf{d} + \textbf{n} 
\label{eq:transmitter}
\end{equation}

where the vector $\textbf{n}$ represents the noise vector formed by $N$ i.i.d. complex Gaussian variables and the matrix $\textbf{H}$ represents the influence of the the frequency selective fadings of the time variant channel on the precoded data. This matrix corresponds to a diagonal matrix of size $N\times{N}$ whose element $ii$ is the channel transfer function of the precoded symbol $i$ (i. e. the element $i$ of the vector $\textbf{s}$. Although in real systems affected by Doppler spread this matrix is not diagonal, for the shake of simplicity we suppose its diagonality. This assumption is realistic in most cases provided that the terminal speed is moderate. Moreover, the employment of the interleaver which separates the carriers corresponding to the same precoded block allows assuming this matrix as diagonal. In this sense, the possible inter channel interference (ICI) corresponding to adjacent carriers would carry data from other precoder blocks and thus can be suitably treated as a penalty in the SNR rather than a loss of the orthogonality of the precoded data.

The received data are then equalized multiplying the vector $\textbf{r}$ with the matrix $\textbf{G}$, which corresponds to a diagonal matrix of size $N\times{N}$ whose element $ii$ is the equalization gain applied to the element $i$ of the received vector, and finally deprecoded with the matrix $\textbf{P}$.  Therefore, the estimated vector data $\hat{\textbf{d}}$ at the system output can be expressed as:
\begin{equation}
\hat{\textbf{d}} = \textbf{P} \, \textbf{G} \, \textbf{r} = \textbf{P} \, \textbf{G} \, \textbf{H} \ \textbf{P} \, \textbf{d} + \textbf{P} \, \textbf{G} \, \textbf{n}
\label{eq:receiver}
\end{equation}

This expression shows that the error in the recovery of the original modulated data can be caused by two different terms. The second term of the expression refers to the error in the received data caused by the noise passed through the equalizer and the deprecoder. On the other hand, the first term corresponds to the loss of orthogonality in the deprecoding process due to the mismatch between the equalizer and the channel. This mismatch can be caused by intercarrier interference, when the matrix $\textbf{H}$ is not diagonal, or by channel estimation errors or the structure of the equalizer itself, when $\textbf{G}$ is not the inverse of $\textbf{H}$.

\section{Statistical characterization of the noise at the deprecoder output}

As stated previously, the second term in (\ref{eq:receiver}) corresponds to the error in the estimated data due to the additive noise passed through the equalizer and the deprecoder at the receiver. In an additive white Gaussian noise (AWGN) channel, the BER of the raw received data for an OFDM system follows an exponentially decay while in a Rayleigh fading channel, this BER follows a linear decay. This different behaviour is produced by the noise amplification at the equalizer of the received carriers which have suffered a deep fading. This fact changes the power and the distribution of the noise at the equalizer output.

The gain introduced by the equalizer in each carrier is a function of the channel attenuation in that carrier. This attenuation can be characterized by a complex circular Gaussian random variable, that is, a random variable whose amplitude has a Rayleigh distribution and its phase a uniform distribution. Thus, the noise component at the equalizer output $v$ is a new random variable formed by the product of the Gaussian channel noise, $n$, and the equalizer, $g$ whose distribution is unknown \textit{a priori}. In order to characterize $v$, we propose to use, in addition to mean and power, the variance of the square of $v$. This measure indicates the variability of the instant noise power distribution amongst the equalized data samples:
\begin{equation}
\mu_v = \mathbf{E} [v] = \mathbf{E} [n] \, \mathbf{E} [g] = 0
\label{eq:mean_v}
\end{equation}
\begin{equation}
\sigma_v^2 = \mathbf{E} [{|v|}^2] =  \mathbf{E}[{|n|}^2] \, \mathbf{E}[{|g|}^2] = \sigma_n^2 \, \sigma_g^2
\label{eq:var_v}
\end{equation}
\begin{equation}
\begin{split}
\mathrm{var} ({|v|}^2) & =  \mathbf{E}\!\left[{({|v|^2-\sigma_v^2})}^2\right] = \mathbf{E}[|v|^4]-\sigma_v^4 \\
   & = \mathbf{E}[|n|^4] \, \mathbf{E}[|g|^4] - \sigma_n^4 \, \sigma_g^4
\end{split}
\label{eq:var_v2}
\end{equation}
where it is assumed that the real and the imaginary part of $n$ and $g$ are both independent. Since $n$ corresponds to a Gaussian random variable, (\ref{eq:var_v2}) can be expressed as:
\begin{equation}
\mathrm{var} ({|v|}^2) = 3\,\sigma_n^4\,\mathbf{E}[|g|^4]-\sigma_n^4 \, \sigma_g^4 = \sigma_n^4\left(3\,\mathbf{E}[|g|^4]-\sigma_g^4\right)
\label{eq:var_v2_gauss_hyp}
\end{equation}

Thus, the equalizer raises the output noise by a factor $\sigma_g^2$ and increases the variability of the instant power distribution of the noise, since   from Jensen's inequality $\mathbf{E}[|g|^4]>\sigma_g^4$. This variability causes the dispersion in the noise power distribution amongst the different carriers in the equalized OFDM symbols, degrading the overall performance of the system.  The noise at the deprecoder output, considering a precoding block of size $N$, can be expressed as:
\begin{equation}
\textbf{w} = \textbf{P} \, \textbf{v} \;\; \Rightarrow \;\; w_j = \displaystyle\sum_{i = 1}^N p_{i,j}\,v_i
\label{eq:w}
\end{equation}
where $\textbf{v}$ and $\textbf{w}$ are  the vectors of equalized noise samples and noise samples at the deprecoder output respectively and $p_{i,j} = \pm 1/\sqrt{N}$ corresponds to the values of the elements of $\textbf{P}$. Characterizing the elements of $\textbf{w}$ with the same moments employed to characterize $v$, it is obtained:
\begin{equation}
\begin{split}
\mu_w & = \mathbf{E} [w] = \mathbf{E} \left[{\displaystyle\sum_{l=1}^N\left({\frac{(-1)^{f(l)}}{\sqrt{N}}\,v_l}\right)}\right]  = \\
   & = \displaystyle\sum_{l=1}^N\left({\frac{(-1)^{f(l)}}{\sqrt{N}}\,\mathbf{E} [v]}\right) = 0
\end{split}
\label{eq:mean_w}
\end{equation}
\begin{equation}
\begin{split}
\sigma_w^2 & = \mathbf{E} [{|w|}^2] = \mathbf{E} \left[{\left|{\displaystyle\sum_{l=1}^N\left({\frac{(-1)^{f(l)}}{\sqrt{N}}\,v_l}\right)}\right|}^2\right]  = \\
   & = \displaystyle\sum_{l=1}^N\left({\frac{l}{N}\,\mathbf{E} [{|v|}^2]}\right) = \sigma_v^2
\end{split}
\label{eq:var_w}
\end{equation}
\begin{equation}
\begin{split}
\mathrm{var} ({|w|}^2) & = \mathbf{E}\!\left[{(|w|^2-\sigma_w^2)}^2\right] = \\ & = \mathbf{E}\!\left[\left({\left|{\displaystyle\sum_{l=1}^N\left({\frac{(-1)^{f(l)}}{\sqrt{N}}\,v_l}\right)}\right|}^2 - \sigma_w^2\right)^2\right]  = \\
   & = \left[{\left|{\displaystyle\sum_{l=1}^N\left({\frac{(-1)^{f(l)}}{\sqrt{N}}\,v_l}\right)}\right|}^4\right] - \sigma_w^4 = \footnotemark \\
& = \frac{1}{N} \mathbf{E} [|v|^4] + 3 \frac{(N-1)}{N} \sigma_w^4-\sigma_w^4 = \\ & = \frac{1}{N} \mathbf{E} [|v|^4] + 3 \frac{(N-1)}{N} \sigma_v^4-\sigma_v^4
\end{split}
\label{eq:var_w2}
\end{equation}

\footnotetext{The expected value of the sum of $N$ independent random variables of null mean and variance $\sigma^2$ to the 4th is:
\begin{flalign*} 
\mathbf{E}[{(X_1 + \ldots +X_N)}^4] & = \mathbf{E}[X_1^4] + 6\mathbf{E}[X_1^2]\mathbf{E}[{(X_2 + \ldots +X_N)}^2] \\
& = \mathbf{E}[X_1^4] + 6(N-1)\sigma^4 + \mathbf{E}[{(X_2 + \ldots +X_N)}^2] = \\
& = N\mathbf{E}[X^4]+6\sigma^4((N-1)+(N-2)+ \ldots + 1) = \\
& = N\mathbf{E}[X^4]+3N(N-1)\sigma^4
\end{flalign*} 
}
with $f(l) = 0, 1$ depending on the element of $\textbf{P}$. Equations (\ref{eq:var_w}) and (\ref{eq:var_w2}) show that the precoding process does not decrease the noise power, but drops the variability of the distribution of that power. In fact the larger the precoding block is, the more uniform is the noise power distribution at the deprecoder output. Thus, if $N\rightarrow{\infty}$, the noise instant power variance tends to:
\begin{equation}
\mathrm{var}({|w|}^2)\approx{2\sigma_v^4}
\end{equation}

\begin{figure}[!t]
\centering
\includegraphics[width=3.6in, clip=true]{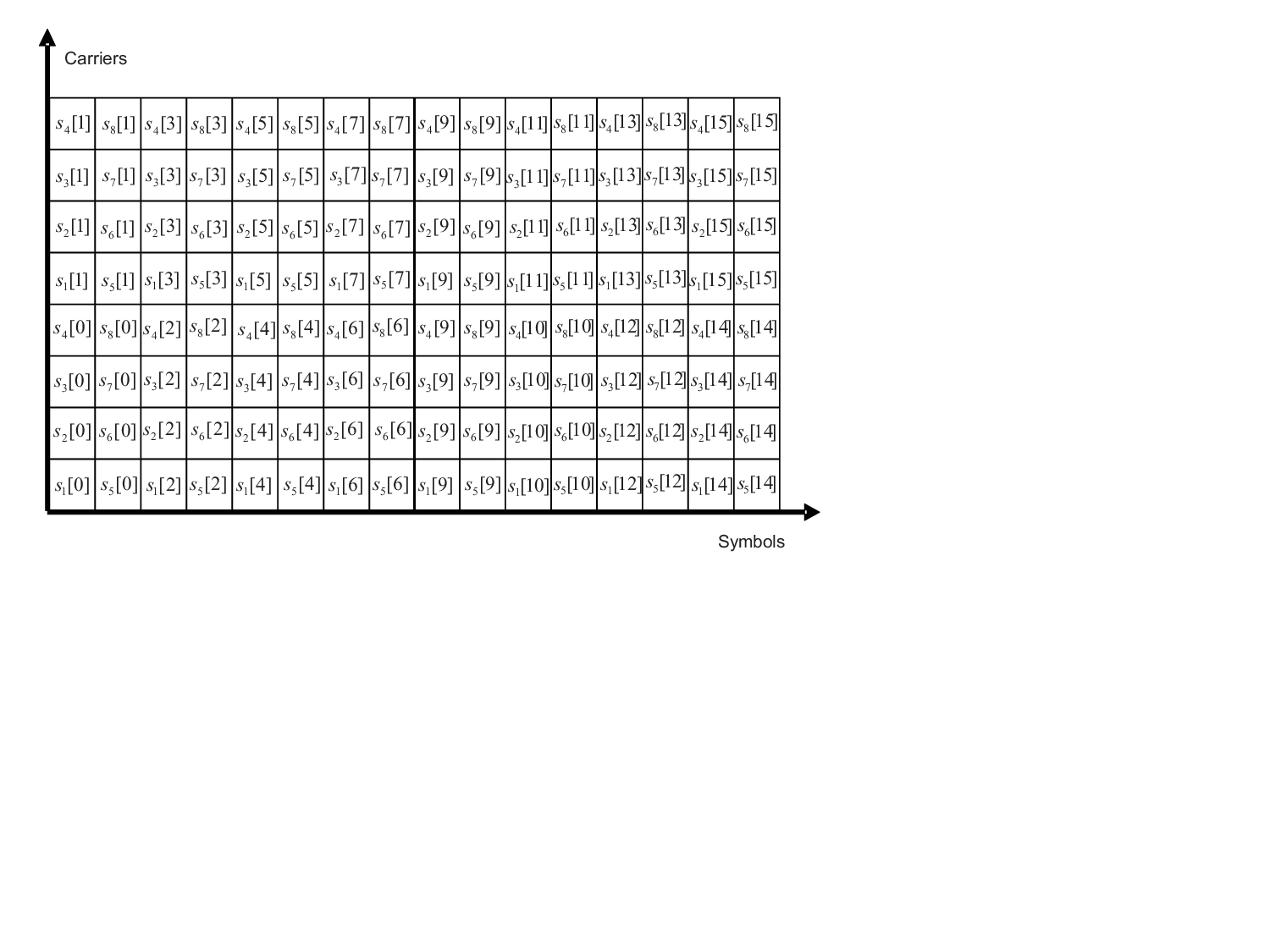}
\caption{Example of a time-frequency interleaving pattern with a precoded block size of 16 symbols and 8 subcarriers per OFDM symbol. Symbols corresponding to the same precoding block are moved apart 4 subcarriers and 2 OFDM symbols.}
\label{fig:interleaver}
\end{figure}

This value corresponds to the variance of the square of a Gaussian random variable with variance $\sigma_v^2$ and null mean, which indicates that the noise at the deprecoder output tends to a Gaussian distribution. In fact, the noise deprecoding process can be seen as the sum of $N$ random variables, which can be approximated by a Gaussian random variable with null mean and variance $\sigma_v^2$. Thus, the higher the size of the precoding block is, the more uniform the noise power distribution will be, which will lead to an increase of the performance of the system.

Finally, the previous results assume the statistical independence of the noise samples at the deprecoder input. In order to achieve this independence, it is necessary the use of a block which interleaves the data corresponding to different precodification blocks. This interleaver has to ensure a separation between samples of the same block of at least the coherence bandwidth (in case the mapping is in the same symbol) or the coherence time (in case the mapping is in different symbols). This block enhances the performance of the proposed system when compared to OFDM-CDM, since in this case the mapping of CDM symbols is performed to consecutive subcarriers of the same OFDM symbol. Fig. \ref{fig:interleaver} shows an example of the operation of the interleaver with a precoded block size of 16 symbols and 8 subcarriers per OFDM symbol. The notation $s_i[j]$ represents the symbol $j$ of the precoded block $i$. In this case, all the symbols corresponding to the same precoding block are separated 4 subcarriers and 2 OFDM symbols.

\section{Equalizer Design}
The first term in (\ref{eq:receiver}) refers to the error in the estimated data due to the impairments introduced by the channel and not compensated by the equalizer. This fact can be caused by the inner structure of the equalizer itself or when the channel estimation is not ideal. If we consider that there are not channel estimation errors and the equalizer at the receiver reverses exactly the channel response (Zero Forcing), this term is null. Nevertheless, such equalizer would amplify the noise at its output to a great degree. While this fact would only affect to a limited number of subcarriers in a system without precodification, the implementation of the precoding scheme would redistribute this noise power amongst all the precoded symbols, leading to a degradation of the system performance. The equalizer cannot hence invert completely the frequency response of the channel and this error will be present in the deprecoded data to an extent.

Assuming absence of noise and ICI and defining the diagonal matrix $\textbf{T}=\textbf{G} \,\textbf{H}$,  whose element $jj$ corresponds to the product of the channel and the equalized gain applied to $s_j$, the estimated data at the deprecoder output can be expressed as:
\begin{equation}
\hat{\textbf{d}} = \textbf{P} \, \textbf{T} \, \textbf{P} \, \textbf{d}
\label{eq:estimated_data}
\end{equation}

Unfolding the matrix products in (\ref{eq:estimated_data}), the element $i$ of $\hat{\textbf{d}}$ can be written as:
\begin{equation}
\begin{split}
\hat{d}_i & = \displaystyle\sum_{j=1}^N p_{i,j}t_{j,j}\left({\displaystyle\sum_{l=1}^N p_{j,l}d_l}\right) = \\
& = \left({\displaystyle\sum_{j=1}^N p_{i,j}t_{j,j}p_{j,i}}\right)d_i + \displaystyle\sum_{\substack{l=1\\l\neq i}}^N\left({\sum_{j=1}^N p_{i,j}t_{j,j}p_{j,l}}\right)d_l
\end{split}
\label{eq:unfolded_estimated_data}
\end{equation}

Since the Hadamard precoding matrix is orthonormal, the self-inner products of its rows must be one and therefore the elementwise products must be $1/N$. In the same way, the inner product of two different rows must be zero, which implies that half the elementwise products must be $1/N$ while the other half are $-1/N$. Thus, it is possible to write (\ref{eq:unfolded_estimated_data}) as:
\begin{equation}
\begin{split}
\hat{d_i} \! = \!\frac {1} {N} \!\! \left({\displaystyle\sum_{j=1}^N t_{j,j}}\!\right)\!d_i+\! \frac {1} {N}\! \displaystyle\sum_{\substack{l=1\\l\neq i}}^N d_l\!\!\left({\!\sum_{\substack{m \ni \\ p_{i,j}p_{j,l} \\ = 1/N}} \!\!\!\! t_{m,m} \! - \!\!\!\!\! \sum_{\substack{n \ni \\ p_{i,j}p_{j,l} \\ = -1/N}} \!\!\!\! t_{n,n}}\!\right)
\end{split}
\label{eq:unfolded_estimated_data_bis}
\end{equation}

The estimation error in noise absence, defined as $\textbf{e} = \hat{\textbf{d}} - \textbf{d}$, can be written as:
\begin{equation}
\textbf{e} = (\textbf{P} \, \textbf{T} \, \textbf{P} - \textbf{I}) \, \textbf{d}
\label{eq:error_vector}
\end{equation}
\begin{equation}
\begin{split}
e_i  = & \frac {1} {N} \left({\displaystyle\sum_{j=1}^N t_{j,j} - 1}\right)d_i \, \, + \\ &  + \frac {1} {N} \displaystyle\sum_{\substack{l=1\\l\neq i}}^N d_l \cdot \left({\sum_{\substack{m \ni \\ p_{i,j}p_{j,l} \\ = 1/N}} \!\!\! t_{m,m} - \!\!\!\! \sum_{\substack{n \ni \\ p_{i,j}p_{j,l} \\ = -1/N}} \!\!\!\! t_{n,n}}\right)
\end{split}
\label{eq:unfolded_estimated_error}
\end{equation}

Assuming the independence amongst the elements of $\textbf{T}$,  which can be achieved with the use of the interleaver, the mean square error can be expressed as:
\begin{equation}
\begin{split}
\!\!\mathbf{E}[ {\left\|{\textbf{e}}\right\|}^2] \! = & \! \underbrace {P_s \! \cdot \mathbf{E}\!\left[\left|{\frac {1} {N} \!\!\left({\displaystyle\sum_{j=1}^N t_{j,j} - 1}\right)}\right|^2\right]}_{\sigma_{dist}^2}+ \\
& + \! \underbrace{\frac {P_s} {N^2} \mathbf{E} \left[ \left| {
\displaystyle\sum_{\substack{l=1\\l\neq i}}^N \! \left({\sum_{\substack{m \ni \\ p_{i,j}p_{j,l} \\ = 1/N}} \!\!\! t_{m,m} \! - \!\!\!\!\! \sum_{\substack{n \ni \\ p_{i,j}p_{j,l} \\ = -1/N}} \!\!\!t_{n,n}}\right)
}\right|^2\right]}_{\sigma_{intf}^2}
\end{split}
\label{eq:mean_square_error}
\end{equation}
where $P_s$ is the signal power. This first term in (\ref{eq:mean_square_error}) encloses the distortion introduced in each symbol, while the second term corresponds to the loss of orthogonality between the symbols of the same precoded block, which can be viewed as an interference in the precoded data. Since the transfer function of the system formed by the channel and the equalizer $t_{i,i}$ can be viewed as a random variable of expected value $E[t]$ and power $E[t^2]$, we can change the matrix notation $t_{i,i}$ by $t_i$. Then, the first term of (\ref{eq:mean_square_error}) can be expressed as: 
\begin{equation}
\begin{split}
\sigma_{dist}^2 & = P_s \!\! \left( \frac {1} {N^2} \mathbf{E}\!\!\left[\!\!{\left( {\displaystyle\sum_{j=1}^N} t_j \! \right)\!\!} ^2 \right] \!\! - \frac {2} {N} \mathbf{E}\!\! \left[ {\displaystyle\sum_{j=1}^N} t_j \right] + 1\right) = \\
& = P_s \left( \frac {1} {N} \mathbf{E}[t^2] + \frac {N-1} {N} \mathbf{E}[t]^2-2\mathbf{E}[t]+1 \right)
\end{split}
\label{eq:var_dist}
\end{equation}

In the same way, if we define the random variable $q$ as the subtraction of two independent random variables $t$, second term of (\ref{eq:mean_square_error}) corresponds to:
\begin{equation}
\begin{split}
\sigma_{intf}^2 & = \frac {P_s} {N^2} \mathbf{E}\left[{\left( 
{\displaystyle\sum_{\substack {l=1 \\ l\neq i}}^N \left({\sum_{m=1}^{N/2} t_m - \sum_{n=1}^{N/2} t_n} \right)\!\!\!} \right)\!\!}^2 \right] = \\
& =  \frac {P_s} {N^2} \mathbf{E}\left[{\left( 
{\displaystyle\sum_{\substack {l=1 \\ l\neq i}}^N \left({\sum_{k=1}^{N/2} q_k} \right)\!\!\!} \right)\!\!}^2 \right] = \\
& = P_s \frac {N-1} {N^2} \, \mathbf{E}\left[{\left( 
{{\sum_{k=1}^{N/2} q_k}\!} \right)\!\!}^2 \right] = P_s \frac {N-1} {2N}\mathbf{E}[q^2]
\end{split}
\label{eq:var_int}
\end{equation}

$\mathbf{E}[q^2]$ can be expressed as a function of $\mathbf{E}[t^2]$ and $\mathbf{E}[t]$:
\begin{equation}
\mathbf{E}[q^2] = \mathbf{E}[(t_1-t_2)^2] = 2\mathbf{E}[t^2]-2\mathbf{E}[t]^2
\label{eq:var_q}
\end{equation}

Substituting (\ref{eq:var_q}) in (\ref{eq:var_int}), it is obtained:
\begin{equation}
\sigma_{intf}^2 = P_s \frac {N-1} {N} (\mathbf{E}[t^2]-\mathbf{E}[t]^2)
\label{eq:var_int2}
\end{equation}

Adding (\ref{eq:var_dist}) and (\ref{eq:var_int2}), we obtain the mean square error as defined in (\ref{eq:mean_square_error}): 
\begin{equation}
\begin{split}
\mathbf{E}[ {\left\|{\textbf{e}}\right\|}^2] & = P_s \! \left(\! \frac {1} {N} \mathbf{E}[t^2] + \! \frac {N\!-\!1} {N} \mathbf{E}[t]^2\! -\!2\mathbf{E}[t]\! + 1 \!\right) + \\
& \quad + P_s \frac {N-1} {N} (\mathbf{E}[t^2]-\mathbf{E}[t]^2) = \\
& = P_s |\mathbf{E}[t]-1|^2
\end{split}
\label{eq:mean_square_error2}
\end{equation}

The result in (\ref{eq:mean_square_error2}) is independent of the precoder block size. Therefore, the interference introduced by the precoding process is compensated by a reduction in the distortion caused by the equalizer, resulting in a net value identical to the distortion introduced by the equalizer in a system without precodification.

The expression of the total mean square error at the deprecoder output is obtained adding the noise power calculated at (\ref{eq:var_w}) to the result of equation (\ref{eq:mean_square_error2}): 

\begin{equation}
\begin{split}
\mathbf{E}[ {|{e}|}^2] & = P_s |\mathbf{E}[t]-1|^2 + \sigma_v^2 \\
& = P_s |\mathbf{E}[g \! \cdot \! h]-1|^2 + \sigma_n^2 \, \sigma_g^2
\label{eq:mean_square_error3}
\end{split}
\end{equation}
where $h$ and $g$ are random variables which represent the channel and the equalizer respectively. As stated previously, if the equalizer reverses completely the channel response, the first term of (\ref{eq:mean_square_error3}) is null. Nevertheless, the term corresponding to the noise contribution to the estimation error would not converge to a finite value since $g$ would be the inverse of a Rayleigh random variable. To prevent this amplification of the noise power, the proposed filter for the signal equalization corresponds to a ZF equalizer whose maximum gain is limited by a clippling threshold. A low threshold will reduce noise amplification at the expense of a distortion raise. The expression of $g$ is:

\begin{equation}
g=\begin{cases}
\dfrac{h^*} {|h|^2}& \text{if $|h| \geq c$},\\
& \\
\dfrac{h^*}{c|h|}& \text{if $|h| < c$}.
\end{cases}
\end{equation}

The minimization of expression (\ref{eq:mean_square_error3}) for each SNR will lead to the optimum clipping threshold $c$ which gives a better performance in terms of BER. This optimum threshold is set heuristically in the next section.

\section{Simulation Results}
In this section, we show the performance of the proposed precoder in the
ITU Vehicular A channel \cite{general:ITU_channels}. This model corresponds to a fading multipath channel with six taps, whose time delays and the variances of the multipath are shown in Table I. The simulated OFDM signal has a bandwidth $B_w$ of 5 MHz and the carrier frequency is $f_c =$ 3.5 GHz. The OFDM symbol period is 102 $\mu$s, with a cyclic prefix of 11 $\mu$s. The number of subcarriers in a symbol is 512, being the subcarrier space 10.9 KHz. The mobile speed is 120 km/h, leading to a Doppler frequency of $f_d = v/\lambda = $ 389 Hz. Since the Doppler frequency is much less than the subcarrier spacing and for the shake of simplicity, the ICI caused by Doppler spread is omitted.

\begin{table}[!t]
\renewcommand{\arraystretch}{1.3}
\caption{ITU Vehicular A Channel Profile}
\label{table:vehicular}
\centering
\begin{tabular}{|c|c|}
\hline
\bfseries Excess Tap & \bfseries Relative \\
\bfseries Delay (ns) & \bfseries Power (dB) \\
\hline
0 & 0\\
\hline
310 & -1\\
\hline
710 & -9\\
\hline
1090 & -10\\
\hline
1730 & -15\\
\hline
2510 & -20\\
\hline
\end{tabular}
\end{table}

The coherence bandwidth estimated from the channel parameters is 430 KHz. Similarly, the coherence time derived from the Doppler frequency is 1.1 ms. Therefore, the interleaver has to separated the data corresponding to the same precoded block at least 40 subcarriers or 11 symbol periods. This choice ensures the independence of the channel suffered by each symbol corresponding to the same precoded block. 

The election of these channel parameters is a representative case of a typical wireless channel. Similar results have been obtained with different values of the channel parameters $f_c$, $v$ and $B_w$ and other channel models (ITU Vehicular B). Changes in these parameters do not affect the overall system performance and only require an adjustment of the interleaver parameters. 

The size of the precoding block is variable and the input data corresponds to QPSK modulated symbols. The mapping of precoded data to subcarriers has been performed considering all the subcarriers in an OFDM symbol of data, so there are not pilot or guard subcarriers. It has been assumed an ideal channel estimation at the receiver. The results have been obtained using Monte-Carlo method.

\begin{figure}[!t]
\centering
\includegraphics[width=3.8in, clip=true]{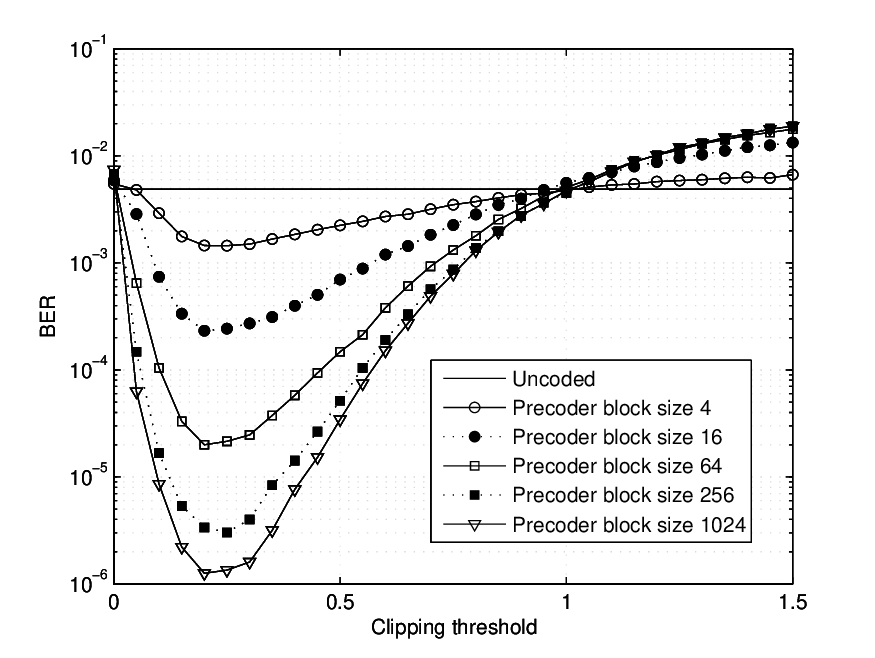}
\caption{BER vs. clipping threshold (SNR = 20 dB)}
\label{fig:20dB}
\end{figure}

\begin{figure}[!t]
\centering
\includegraphics[width=3.8in, clip=true]{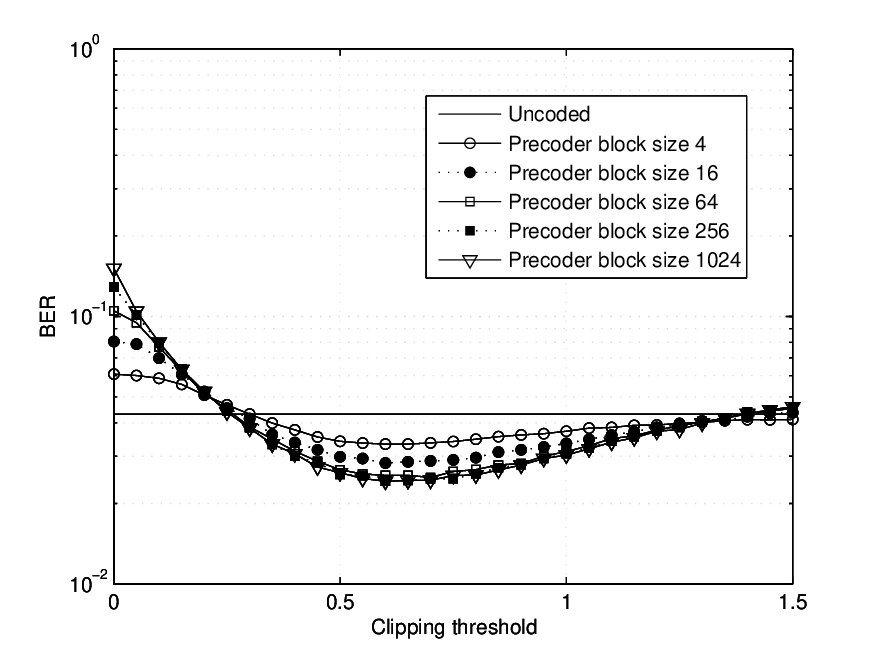}
\caption{BER vs. clipping threshold (SNR = 10 dB)}
\label{fig:10dB}
\end{figure}

Fig. \ref{fig:20dB} and \ref{fig:10dB} show the BER obtained with different precoder block sizes as a function of the clipping threshold, $c$, used in the equalizer for a SNR of 20 and 10 dB respectively. The optimum $c$ falls with increasing SNR, which is due to the trade off between the noise amplification at the equalizer output and the distortion introduced to the transmitted signal. The higher $c$ is, the lesser the noise power at the output is and the higher the distortion is. Therefore, for high SNR the estimation error produced by the distortion is more significant than the error caused by the noise and the equalizer tries to minimize it. On the contrary, when the SNR decays, the equalizer tries to minimize the noise amplification rising the clipping threshold. As expected, this optimum clipping threshold is independent of the precoder block size since the noise power and the distortion do not depend of the precoding process. The huge range of clipping thresholds which beats the performance of a non precoded system indicates the robustness of the proposed equalizer. 

\begin{figure}[!t]
\centering
\includegraphics[width=3.8in, clip=true]{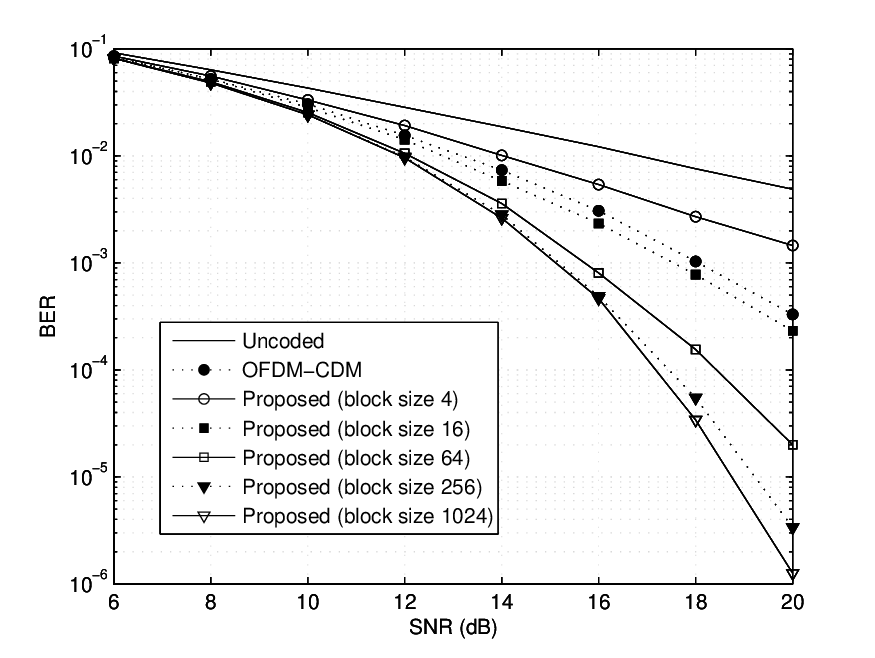}
\caption{BER comparison for different precoder block sizes assuming optimum clipping threshold}
\label{fig:BER}
\end{figure}

In Fig. \ref{fig:BER}, the BER of the proposed system for different precoding block sizes is compared to those obtained with an uncoded OFDM system and a typical OFDM-CDM system. It is supposed that the optimum clipping threshold is set in the equalizer in the considered range of SNR. The simulated OFDM-CDM system consists of a precoder whose block size is 512 symbols and which is also based on a Hadamard matrix. The data corresponding to the each precoded block are mapped to consecutive subcarriers, hence matching each CDM block to an OFDM symbol. The obtained results improve for all the block sizes the BER of the system without precoder. Likewise, the proposed system improves the results of the OFDM-CDM system when the precoded block size is 16 or greater. In this sense, the proposed system with precoding block sizes between 16 and 512 symbols would overcome the performance of the OFDM-CDM system with a lower computational load. It can also be seen that the performance of the proposed system tends to an exponential decay for increasing sizes of the precoding block. This effect confirms that the noise at the precoder output follows a gaussian distribution.

\begin{figure}[!t]
\centering
\includegraphics[width=3.8in, clip=true]{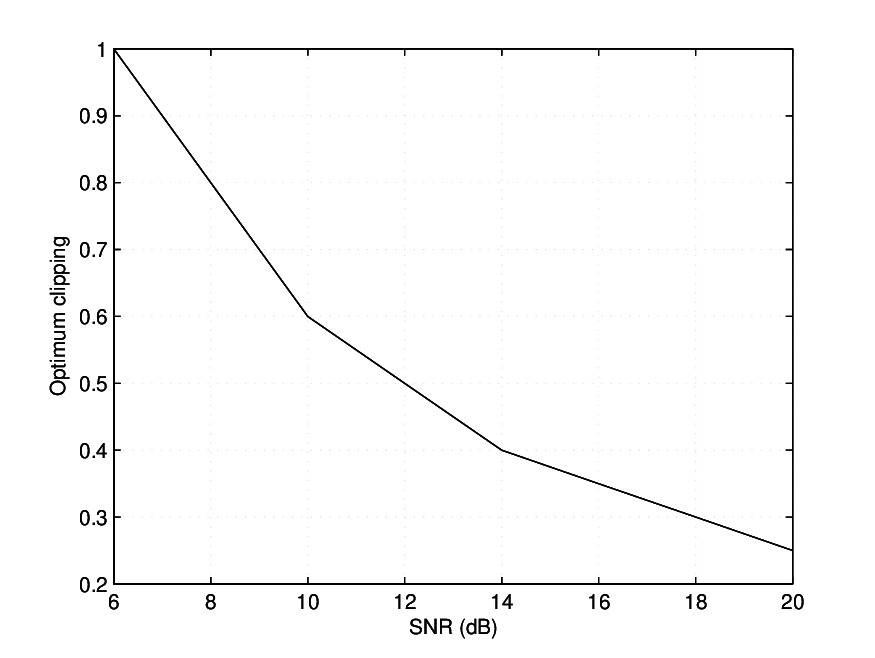}
\caption{Optimum clipping threshold of the proposed equalizer as a function of the SNR}
\label{fig:clipping}
\end{figure}

Fig. \ref{fig:clipping} illustrates the optimum threshold in the equalizer as a function of the SNR. This optimal threshold decreases for high SNR since the equalizer tends to reverse the channel more accurately in order to minimize the distortion in the estimated signal.

\begin{figure}[!t]
\centering
\includegraphics[width=3.8in, clip=true]{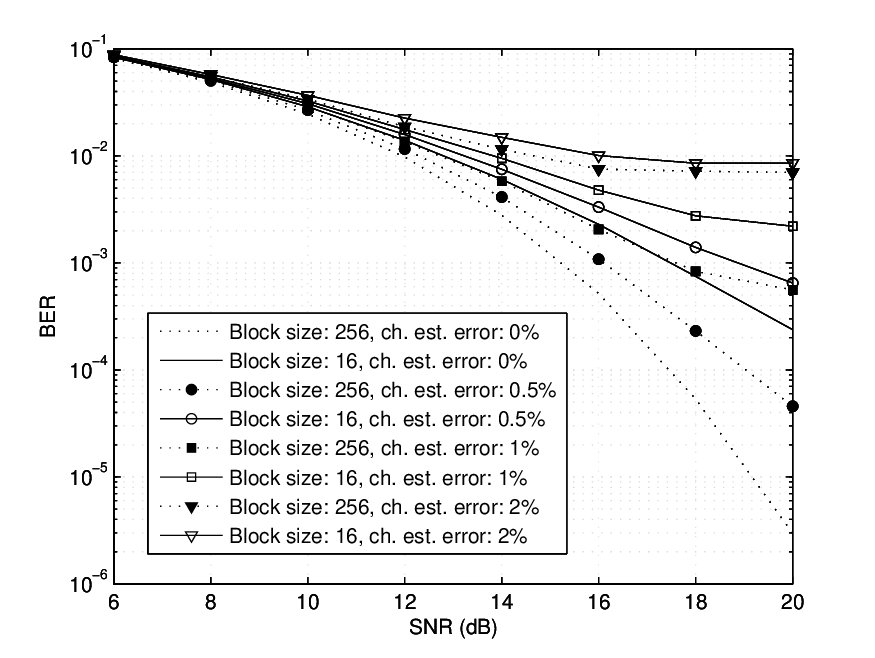}
\caption{BER comparison for different precoder block sizes and channel estimation errors}
\label{fig:CSI_error}
\end{figure}

Finally, figure \ref{fig:CSI_error} shows the performance of the proposed systems in the presence of impairments in the channel estimation. These impairments have been modelled as a complex circular Gaussian random variable added to the ideal channel value which employs the equalizer. The error is defined as the mean square of this random variable, since the mean square of the channel is equal to 1. The performance of the system has been derived for precoded block sizes of 16 and 256 symbols. In both cases, the penalty in the BER is about 1dB when the estimation error is 0.5\% and 2dB when the error is 1\%.

\section{Conclusion}
A channel independent precoder is proposed to improve the performance of OFDM systems over fading channels. The precoder has a variable block size and is employed together with an interleaver in order to ensure the channel diversity of the precoded data. The proposed precoder does not require any prior knowledge of the channel conditions at the transmitter, making the system suitable for broadcast applications. Unlike other previous solutions, the block size of the proposed precoder could be different from the FFT size, achieving an independence between the precoding block and the specific parameters of the OFDM symbol employed in the system. The equalization at the receiver is done with a zero forcing filter whose maximum gain is limited with a clipping threshold in order to minimize the noise power transfer amongst precoded data. Simulation results show that the proposed system outperforms the BER obtained in a non-precoded OFDM sytem as well as in a typical OFDM-CDM system.

\ifCLASSOPTIONcaptionsoff
  \newpage
\fi

\bibliographystyle{IEEEtran}
\bibliography{./IEEEabrv,./bibliografia}

\begin{thebibliography}{10}
\providecommand{\url}[1]{#1}
\csname url@samestyle\endcsname
\providecommand{\newblock}{\relax}
\providecommand{\bibinfo}[2]{#2}
\providecommand{\BIBentrySTDinterwordspacing}{\spaceskip=0pt\relax}
\providecommand{\BIBentryALTinterwordstretchfactor}{4}
\providecommand{\BIBentryALTinterwordspacing}{\spaceskip=\fontdimen2\font plus
\BIBentryALTinterwordstretchfactor\fontdimen3\font minus
  \fontdimen4\font\relax}
\providecommand{\BIBforeignlanguage}[2]{{%
\expandafter\ifx\csname l@#1\endcsname\relax
\typeout{** WARNING: IEEEtran.bst: No hyphenation pattern has been}%
\typeout{** loaded for the language `#1'. Using the pattern for}%
\typeout{** the default language instead.}%
\else
\language=\csname l@#1\endcsname
\fi
#2}}
\providecommand{\BIBdecl}{\relax}
\BIBdecl

\bibitem{general:xDSL}
J.~S. Chow, J.~C. Tu, and J.~M. Cioffi, ``A discrete multitone transceiver
  system for {HDSL} applications,'' \emph{{IEEE} J. Sel. Areas Commun.},
  vol.~9, pp. 895--908, Aug. 1991.

\bibitem{general:DAB}
\emph{Digital Audio Broadcasting {(DAB)} to mobile, portable and fixed
  receivers}, ETSI Std. ETS 300 401, 1994.

\bibitem{general:DVB-T}
\emph{Framing structure, channel coding and modulation for digital terrestrial
  television}, ETSI Std. EN 300 744, 1994.

\bibitem{general:DVB-H}
\emph{Transmission system for handheld terminals}, ETSI Std. EN 302 304, 1994.

\bibitem{general:802.11}
\emph{{IEEE} {802.11g-2003}: Further Higher-Speed Physical Layer Extension in
  the 2.4 GHz Band}, IEEE Std., 2003.

\bibitem{general:802.16}
\emph{{IEEE} {Std 802.16e-2005}: Amendment to {IEEE} {Std} {802.16}}, IEEE
  Std., 2005.

\bibitem{general:LTE}
H.~Ekstrom, A.~Furuskar, J.~Karlsson, M.~Meyer, S.~Parkvall, J.~Torsner, and
  M.~Wahlqvist, ``Technical solutions for the {3G} long-term evolution,''
  \emph{{IEEE} Commun. Mag.}, vol.~44, pp. 38--45, Mar. 2006.

\bibitem{precoding:Wang1}
Z.~Wang and G.~B. Giannakis, ``Linearly precoded or coded {OFDM} against
  wireless channel fades?'' in \emph{Third {IEEE} Workshop Signal Process. Adv.
  Wireless Commun.}, Taoyuan, Taiwan, Mar. 2001.

\bibitem{precoding:Ding1}
Y.~Ding, T.~N. Davidson, Z.~Luo, and K.~M. Wong, ``Minimum {BER} block
  precoders for zero-forcing equalizations,'' \emph{{IEEE} Trans. Signal
  Process.}, vol.~51, pp. 2410--2423, Sep. 2003.

\bibitem{precoding:Scaglione1}
A.~Scaglione, S.~Barbarossa, and G.~B. Giannakis, ``Filterbank transceivers
  optimizing information rate in block transmissions over dispersive
  channels,'' \emph{{IEEE} Trans. Inf. Theory}, vol.~45, pp. 1019--1032, Apr.
  1999.

\bibitem{precoding:Liu1}
Z.~Liu, Y.~Xin, and G.~B. Giannakis, ``Linear constellation precoding for
  {OFDM} with maximum multipath diversity and coding gains,'' \emph{{IEEE}
  Trans. Commun.}, vol.~51, pp. 416--427, Mar. 2003.

\bibitem{precoding:Rong1}
Y.~Rong, S.~A. Vorobyov, , and A.~B. Gershman, ``Linear block precoding for
  {OFDM} systems based on maximization of mean cutoff rate,'' \emph{{IEEE}
  Trans. Signal Process.}, vol.~53, pp. 4691--4696, Dec. 2005.

\bibitem{precoding:Lin1}
Y.~P. Lin and S.~F. Phoong, ``{BER} minimized {OFDM} systems with channel
  independent precoders,'' \emph{{IEEE} Trans. Signal Process.}, vol.~51, pp.
  2369--2380, Sep. 2003.

\bibitem{precoding:Kaiser}
S.~Kaiser, ``{OFDM} code division multiplexing in fading channels,''
  \emph{{IEEE} Trans. Commun.}, vol.~50, pp. 1266--1273, Aug. 2002.

\bibitem{precoding:book_precoding}
A.~F. Molisch, \emph{Wideband Wireless Digital Communications}.\hskip 1em plus
  0.5em minus 0.4em\relax Prentice Hall, 2000.

\bibitem{MC-CDMA:Yee1}
N.~Yee, J.~Linnartz, and G.~Fettweis, ``Multi-carrier {CDMA} in indoor wireless
  radio networks,'' in \emph{{IEEE} Personal Indoor and Mobile Radio
  Communications (PIMRC) Int. Conference}, Yokohama, Japan, Sep. 1993.

\bibitem{MC-CDMA:Chin1}
A.~S. Madhukumar and F.~Chin, ``Residue number system-based multicarrier {CDMA}
  system for high-speed broadband wireless access frequency-time spreading,''
  \emph{{IEEE} Trans. Broadcast.}, vol.~48, pp. 46--52, Mar. 2002.

\bibitem{MC-CDMA:Zheng1}
K.~Zheng, G.~Zeng, and W.~Wang, ``Performance analysis for {OFDM-CDMA} with
  joint frequency-time spreading,'' \emph{{IEEE} Trans. Broadcast.}, vol.~51,
  pp. 144--148, Mar. 2005.

\bibitem{general:ITU_channels}
``Guidelines for the evaluation of radio transmission technologies for
  {IMT-2000},'' Recommendation ITU-R M.1225, 1997.

\end{thebibliography}
\end{document}